\documentclass[preprint,showpacs,preprintnumbers,amsmath,amssymb]{revtex4}
\usepackage{graphics}
\usepackage{bm}

\begin{document}

\title{Negative Energy: From Lamb Shift to Entanglement}

\author{Shou-Liang Bu}
\email{shouliang@jyu.edu.cn}

\affiliation{School of Physics and Optical Information Technology,
Jiaying University, 514015 MeiZhou, China}

\date{\today}

\begin{abstract}
``Negative energy'' has been one of the most enduring puzzles in
quantum theory, whereas the present work reveals that it actually
plays a central role in clarifying various controversies of
quantum theory. The basic idea is contained in a hypothesis on
negative energy, and it is shown that the idea: (1)is compatible
with both relativistic quantum mechanics and known experimental
results; (2)helps to clarify the essence of matter waves, and
therefore better understand the reality of the wave function, the
so-called `wave-packet reduction' occurring in quantum
measurement, and the ghost like correlations between entangled
systems; (3)is helpful for distinguishing the vacuum from the
ground state of the quantized field, and may supply a possible way
for removing the deep-rooted infinities in quantum field theory.
The vacuum energy density of the electromagnetic field is
calculated here as an example. By employing the same idea, the
Lamb-Shift is recalculated in a different way from conventional
renormalization method, yet the same result as Bethe's can be
definitely obtained.
\end{abstract}


\maketitle

\section{Introduction}
Controversies continuously associate with quantum mechanics since
its foundation in the early twentieth century, ranging from the
reality of the wave function, the so-called `wave-packet
reduction' occurring in quantum measurement, the series of
infinities emerging in quantum field theory, to the latest
nonlocal correlations between quantum entangled systems.

For instance, more and more theoretical and experimental works
demonstrate that entangled systems may violate the separability
principle\cite{Buscemi,Christensen,Popescu}, i.e., two dynamically
independent systems possess their own separate state, advocated by
Einstein et al. One cannot but ask how the `nonlocal correlations'
between entangled systems are established, by considering that
there exist bounds on the speed of transmission of signals or
physical effects as specified by special relativity.

The problem of negative energies is another most enduring puzzle
in quantum theory. It is well known that relativistic quantum
mechanics contains negative energy states as well as familiar
positive ones. In order to ensure the completeness of state space,
these negative energy states can not be simply discarded, however
if people directly accept these negative energy states, one
problem then arises, namely an electron in a positive energy state
should be able to emit a photon and then make a transition to a
negative energy state; moreover the process could even evolve into
a disastrous cascades by continuously emitting an infinite amount
of photons. Historically, in order to solve this problem, Dirac
had imagined that all the states of negative energy are occupied
and the Pauli exclusive principle thus keeps positive energy
electrons from making transitions to negative energy
states\cite{Dirac}. However, this theory faces a number of new and
immediate difficulties such as what about all that change and
negative energy, where is electric field of the ubiquitous
negative energy electron, why is there an asymmetry in the vacuum
between negative and positive energy, and so forth.

In the present work, it is demonstrated that the negative energy
is not only compatible with the theory, but also helpful for
solving various puzzles and controversies in quantum theory.
Briefly speaking, once the existence of negative energy (and
matter) is identified, we can more deeply understand the reality
of the matter waves, the relations between particle and wave, and
the essence of the so-called `wavepacket reduction'; and we can
further reveal the secrets hidden in the double-slit interference
and the entangled systems. Moreover, by using of the property of
matter with negative energies, the infinity of vacuum energy
density emerged in the traditional quantum field theory can be
easily removed, and the Lamb-shift can also be correctly derived
in a completely different way from conventional renormalization
method. Additionally, in the light of the idea presented here, we
may rethink about the nature of the vacuum, and distinguish the
vacuum from the ground state of the quantized field, and this may
further point out a possible way for removing other infinities
rooted in the traditional quantum field theory.

The basic idea is contained in the following hypothesis: {\it{For
each kind of particle with positive energy, there always exists a
kind of negative energy particle which owns completely opposite
intrinsic properties, and the physical laws both kinds of particle
obey have the same covariant form.}}

In particular, the intrinsic property includes the rest mass of
the particle.

The structure of the paper is as follows. First, in section 2, the
mathematical descriptions of negative as well as the positive
energy matter are presented. Then in section 3 and 4, the vacuum
energy density and the Lamb-Shift is recalculated in the light of
the present idea. In section 5, the essence of matter waves,
relations between particle and wave, the so-called `wave packet
reduction', etc. are deeply explored. Section 6 focuses on the
superposition state, in particular, the double-slit experiment and
the entangled systems are analyzed. Finally, in section 7, some
discussions about the vacuum versus the ground state of the
quantized field are given, and it is illustrated that the
traditional quantum field theory is actually equivalent to an
extreme case: the vacuum has an infinite formal temperature.

\section{The Mathematical Descriptions}
Here, a mathematical description for both positive and negative
energy field is developed, for specification, we take the
electromagnetic field and the Dirac field as typical examples.
\\
{\it{The Electromagnetic Field}} --- Let $A^{(\pm)\mu}(x)$
describe the photons fields with superscripts $(\pm)$ denoting
positive and negative energy, according to the hypothesis, the
covariant Maxwell equations for free electromagnetic field are
\begin{equation}\label{Max-eqP}
\Box A^{(\pm)\mu}(x) = 0.
\end{equation}
To derive Eq.~(\ref{Max-eqP}), let's set the Lagrangian density of
positive and negative energy field as follows,
\begin{equation}\label{Max-LagP}
\mathcal{L}^{(\pm)}  =
\mp\frac{1}{2}\left(\partial_{\mu}A^{(\pm)}_{\mu}(x)\right)\left(\partial^{\nu}A^{(\pm)\nu}(x)\right),
\end{equation}
with the constraint $\partial_{\mu}A^{(\pm)\mu}(x) = 0$. It should
be noted that Lagrangian density proposed here is not necessary
the only possible choice for mathematical realization of the
hypothesis.

In order to quantize the electromagnetic fields, one takes
\begin{equation}\label{Max-quanP}
A^{(\pm)\mu}(x) = \sum_{r\mathbf{k}}\sqrt{\frac{\hbar
c^2}{2V\omega_{\mathbf{k}}}}\left[\varepsilon_{r}^{\mu}(\mathbf{k})a^{(\pm)}_{r}(\mathbf{k})e^{-ikx}
+
\varepsilon_{r}^{\mu}(\mathbf{k})a_{r}^{(\pm)\dag}(\mathbf{k})e^{ikx}\right],
\end{equation}
with $r = 0, \cdots, 3$, $k^{0} = \omega_{\mathbf{k}}/c =
|\mathbf{k}|$, and the vectors $\varepsilon_{r}^{\mu}(\mathbf{k})$
describe four linearly independent polarization states. The
commutation relations are
\begin{eqnarray}\label{Max-comP}
\left[a^{(\pm)}_{r}(\mathbf{k}),
a_{s}^{(\pm)\dag}(\mathbf{k'})\right] &=&
\zeta_{r}\delta_{rs}\delta_{\mathbf{k}\mathbf{k'}}, \nonumber\\
\left[a^{(\pm)}_{r}(\mathbf{k}), a^{(\pm)}_{s}(\mathbf{k'})\right]
&=& \left[a_{r}^{(\pm)\dag}(\mathbf{k}),
a_{s}^{(\pm)\dag}(\mathbf{k'})\right] = 0,
\end{eqnarray}
here, $\zeta_{r} = 1$ for $r = 1, 2, 3$, and $-1$ for $r = 0$.

From Eq.~(\ref{Max-eqP}) - (\ref{Max-quanP}), the Hamiltonian
operator of the positive and negative energy fields are given by
\begin{eqnarray}\label{Max-HamP}
\mathcal{H}^{(\pm)} &=& \int d^{3}\mathbf{x}\left[
\pi^{(\pm)\mu}(x)\dot{A}^{(\pm)}_{\mu}(x) - \mathcal{L}^{(\pm)}
\right]
\nonumber\\
&=& \sum_{r\mathbf{k}}(\pm\hbar\omega_{\mathbf{k}})
\zeta_{r}\left[a_{r}^{(\pm)\dag}(\mathbf{k})a^{(\pm)}_{r}(\mathbf{k})
+ \frac{1}{2} \right],
\end{eqnarray}
here $\pi^{(\pm)\mu}(x) =
\frac{\partial\mathcal{L}^{(\pm)}}{\partial\dot{A}^{(\pm)}_{\mu}}
= -\frac{1}{c^{2}}\dot{A}^{(\pm)}_{\mu}(x)$. Other observables
such as momentum, angular momentum etc. can be similarly obtained.

Eq.~(\ref{Max-HamP}) shows that, as
$a_{r}^{(+)\dag}(\mathbf{k})a_{r}^{(+)}(\mathbf{k})$ being the
number operator of the positive energy photon
$(+\hbar\omega_{\mathbf{k}})$,
$a_{r}^{(-)\dag}(\mathbf{k})a_{r}^{(-)}(\mathbf{k})$ denotes the
number operator of the negative energy photon
$(-\hbar\omega_{\mathbf{k}})$. Let
$\{|n_{r}^{(+)}(\mathbf{k})\rangle, (n_{r}^{(+)}(\mathbf{k}) =
0,1,2,\ldots)\}$ denote the orthonormal eigenvectors of
$a_{r}^{(+)\dag}(\mathbf{k})a_{r}^{(+)}(\mathbf{k})$, then it is
familiarly known that
$a_{r}^{(+)\dag}(\mathbf{k})|n_{r}^{(+)}(\mathbf{k})\rangle =
\sqrt{n_{r}^{(+)}(\mathbf{k})+1}|n_{r}^{(+)}(\mathbf{k})+1\rangle$
and $a_{r}^{(+)}(\mathbf{k})|n_{r}^{(+)}(\mathbf{k})\rangle =
\sqrt{n_{r}^{(+)}(\mathbf{k})}|n_{r}^{(+)}(\mathbf{k})-1\rangle$
(with $n_{r}^{(+)}(\mathbf{k})\neq 0 $). Analogously, for negative
energy field, let $\{|n_{r}^{(-)}(\mathbf{k})\rangle,
(n_{r}^{(-)}(\mathbf{k}) = 0,1,2,\ldots)\}$ denote the orthonormal
eigenvectors of
$a_{r}^{(-)\dag}(\mathbf{k})a_{r}^{(-)}(\mathbf{k})$, i.e.,
$a_{r}^{(-)\dag}(\mathbf{k})a_{r}^{(-)}(\mathbf{k})|n_{r}^{(-)}(\mathbf{k})\rangle
= n_{r}^{(-)}(\mathbf{k})|n_{r}^{(-)}(\mathbf{k})\rangle$, then by
using Eq.~(\ref{Max-comP}) it can be easily proved that
$a_{r}^{(-)\dag}(\mathbf{k})|n_{r}^{(-)}(\mathbf{k})\rangle =
\sqrt{n_{r}^{(-)}(\mathbf{k})+1}|n_{r}^{(-)}(\mathbf{k})+1\rangle$
and $a_{r}^{(-)}(\mathbf{k})|n_{r}^{(-)}(\mathbf{k})\rangle =
\sqrt{n_{r}^{(-)}(\mathbf{k})}|n_{r}^{(-)}(\mathbf{k})-1\rangle$
(with $n_{r}^{(-)}(\mathbf{k})\neq 0 $).

From Eq.~(\ref{Max-HamP}),
$\mathcal{H}^{(+)}|n_{r}^{(+)}(\mathbf{k})\rangle =
+\hbar\omega_{\mathbf{k}}(n_{r}^{(+)}(\mathbf{k})
+\frac{1}{2})|n_{r}^{(+)}(\mathbf{k})\rangle$, while
$\mathcal{H}^{(-)}|n_{r}^{(-)}(\mathbf{k})\rangle =
-\hbar\omega_{\mathbf{k}}(n_{r}^{(-)}(\mathbf{k})
+\frac{1}{2})|n_{r}^{(-)}(\mathbf{k})\rangle$. For the positive
energy field, $\{+\hbar\omega_{\mathbf{k}}(n_{r}^{(+)}(\mathbf{k})
+\frac{1}{2})\}$ ($n_{r}^{(+)}(\mathbf{k}) \geq 1$) are the energy
of the $n$th excited state, while
$\{-\hbar\omega_{\mathbf{k}}(n_{r}^{(-)}(\mathbf{k})
+\frac{1}{2})\}$ ($n_{r}^{(-)}(\mathbf{k}) \geq 1$) the energy of
the $n$th excited state of the negative energy field. In the case
of positive energy, the height of a state's energy is considered
to be determined by the amount of the number of photons
$(+\hbar\omega_{\mathbf{k}})$, symmetrically, the height of a
state of negative energy field should also be determined by the
amount of the number of photons $(-\hbar\omega_{\mathbf{k}})$.
Therefore, just like $+(\mu_{r}^{(+)}(\mathbf{k})
+\frac{1}{2})\hbar\omega_{\mathbf{k}}$ is a higher level than
$+(\nu_{r}^{(+)}(\mathbf{k})
+\frac{1}{2})\hbar\omega_{\mathbf{k}}$ when
$\mu_{r}^{(+)}(\mathbf{k})>\nu_{r}^{(+)}(\mathbf{k})$,
$-(\mu_{r}^{(-)}(\mathbf{k})
+\frac{1}{2})\hbar\omega_{\mathbf{k}}$ is also a higher level than
$-(\nu_{r}^{(-)}(\mathbf{k})
+\frac{1}{2})\hbar\omega_{\mathbf{k}}$ for negative energy field
if $\mu_{r}^{(-)}(\mathbf{k})>\nu_{r}^{(-)}(\mathbf{k})$. The
statements discussed here can be easily generalized to other
quantized fields.
\\
{\it{The Dirac Field}} --- In terms of the hypothesis, the Dirac
equation for positive and negative rest mass $\pm m$ has the same
form
\begin{equation}\label{Dir-eqP}
\left(i\hbar \gamma^{\mu}\partial_{\mu} - mc\right)\psi^{(\pm)}(x)
= 0,
\end{equation}
with $m>0$ and $\gamma^{\mu}\ (\mu = 0, 1, 2, 3)$ are Dirac $4
\times 4$ matrices. To derive the Dirac equation ~(\ref{Dir-eqP}),
set the Lagrangian density of positive and negative energy field
as
\begin{equation}\label{Dir-LagP}
\mathcal{L}^{(\pm)}  = \pm c\bar{\psi}^{(\pm)}(x)\left( i\hbar
\gamma^{\mu}\partial_{\mu} - mc \right)\psi^{(\pm)}(x),
\end{equation}
with $\bar{\psi}^{(\pm)}(x) = \psi^{(\pm)\dag}(x)\gamma^{0}$.

By expanding it in terms of the complete set of plane wave
solutions, the Dirac field is quantized,
\begin{equation}\label{Dir-quanP}
\psi^{(\pm)}(x) =
\sum_{r\mathbf{p}}\sqrt{\frac{mc^2}{VE_{\mathbf{p}}}}\left[
c^{(\pm)}_r(\mathbf{p})u_r(\mathbf{p})e^{-ipx/\hbar} +
d_r^{(\pm)\dag}(\mathbf{p})v_r(\mathbf{p})e^{ipx/\hbar} \right],
\end{equation}
with $E_{\mathbf{p}} = cp_0 = +\sqrt{m^2c^4+c^2\mathbf{p}^2}$, and
the commutation relation:
\begin{eqnarray}\label{Dir-comP}
\left[c^{(\pm)}_{r}(\mathbf{p}),
c_{s}^{(\pm)\dag}(\mathbf{p'})\right] &=&
\left[d^{(\pm)}_{r}(\mathbf{p}),
d_{s}^{(\pm)\dag}(\mathbf{p'})\right] =
\delta_{rs}\delta_{\mathbf{p}\mathbf{p'}}, \nonumber\\
\left[c^{(\pm)}_{r}(\mathbf{p}), c^{(\pm)}_{s}(\mathbf{p'})\right]
&=& \left[d_{r}^{(\pm)\dag}(\mathbf{p}),
d_{s}^{(\pm)\dag}(\mathbf{p'})\right] = 0.
\end{eqnarray}
The Hamiltonian of the Dirac field, from Eq.~(\ref{Dir-eqP})
-~(\ref{Dir-comP}) is given by
\begin{eqnarray}\label{Dir-HamP}
\mathcal{H}^{(\pm)} &=& \int
d^{3}\mathbf{x}\bar{\psi}^{(\pm)}(x)\left[ -i\hbar
c\gamma^{j}\partial_{j}+mc^2 \right]\psi^{(\pm)}(x)
\nonumber\\
&=& \int
d^{3}\mathbf{x}\left\{\pi^{(\pm)}_j(x)\dot{\psi}^{(\pm)}_j(x)
+\bar{\pi}^{(\pm)}_j(x)\bar{\psi}^{(\pm)}_j(x)-\mathcal{L}^{(\pm)}
\right\}\nonumber\\
&=& \sum_{r\mathbf{p}}(\pm
E_{\mathbf{p}})\left[c_{r}^{(\pm)\dag}(\mathbf{p})c^{(\pm)}_{r}(\mathbf{p})
+d_{r}^{(\pm)\dag}(\mathbf{p})d^{(\pm)}_{r}(\mathbf{p}) -1
\right],
\end{eqnarray}
here
$\pi^{(\pm)}_{\alpha}(x)=\frac{\partial\mathcal{L}^{(\pm)}}{\partial
\dot{\psi}^{(\pm)}_{\alpha}(x)}=i\hbar
\psi_{\alpha}^{(\pm)\dag}(x)$,
$\bar{\pi}^{(\pm)}_\alpha(x)=\frac{\partial\mathcal{L}^{(\pm)}}{\partial
\dot{\bar{\psi}}^{(\pm)}_{\alpha}}=0$. One notes that, for the
ground state, there is an infinite energy for positive or negative
field alone. In principle, it is not difficult to derive the
mathematical forms of other physical quantities, however, this is
not the main goal of the present work.

Eq.~(\ref{Dir-HamP}) tells that, if let
$c_{r}^{(+)\dag}(\mathbf{p})c^{(+)}_{r}(\mathbf{p})$ and
$d_{r}^{(+)\dag}(\mathbf{p})d^{(+)}_{r}(\mathbf{p})$ denote the
number operator of the particle $(+m,\ -e)$ and $(+m,\ +e)$,
respectively, $c_{r}^{(-)\dag}(\mathbf{p})c^{(-)}_{r}(\mathbf{p})$
and $d_{r}^{(-)\dag}(\mathbf{p})d^{(-)}_{r}(\mathbf{p})$ then
denote the number operator of the particle $(-m,\ -e)$ and $(-m,\
+e)$, respectively. Other observable quantities such as momentum
etc. can be similarly derived.

Though we have just investigated the Maxwell and Dirac field here,
the same idea can also be applied to other cases such as
Klein-Gordon field and gravitation field, etc.

Now, let's reconsider the fictitious transition between positive
and negative energy states. In previous considerations, it is
taken for granted that the same one positive mass particle has
equally the positive and negative energy states, and this
assumption further leads to the unreal transitions between
positive and negative energy states. As seen from
Eq.~(\ref{Max-eqP}), (\ref{Max-HamP}), (\ref{Dir-eqP}) and
(\ref{Dir-HamP}), different kinds of particle may obey the same
covariant equation, whereas the positive and negative energy
states are respectively attributed to particles with positive and
negative energy. As is known for positive energy states, the
energy level of a state is considered to be higher, if this state
contains more number of quantum. Thus, the lowest energy state is
$\left\{|+\frac{1}{2}\hbar \omega_{\mathbf{k}}\rangle\right\}$,
and the excited states are $\left\{|n\hbar
\omega_{\mathbf{k}}+\frac{1}{2}\hbar
\omega_{\mathbf{k}}\rangle\right\}$ with $n\geq1$. It is similar
for negative energy states, if a state contains more number of
quantum, its energy level is more high. Therefore, the lowest
energy state is $\left\{|-\frac{1}{2}\hbar
\omega_{\mathbf{k}}\rangle\right\}$, and the excited states are
$\left\{|-n\hbar \omega_{\mathbf{k}}-\frac{1}{2}\hbar
\omega_{\mathbf{k}}\rangle\right\}\ (n\geq1) $, respectively.
Negative energy particle jumping from the ground state to the
excited state, or from the state $\left\{|-n_1\hbar
\omega_{\mathbf{k}}-\frac{1}{2}\hbar
\omega_{\mathbf{k}}\rangle\right\}$ to $\left\{|-n_2\hbar
\omega_{\mathbf{k}}-\frac{1}{2}\hbar
\omega_{\mathbf{k}}\rangle\right\}$ with $n_1<n_2$, needs to
absorb one ore more negative energy photon(s) with energy
$\left(-\hbar \omega_{\mathbf{k}}\right)$ rather than emit
positive energy photon(s) with energy $\left(+\hbar
\omega_{\mathbf{k}}\right)$, while the fictitious transition
between positive and negative energy states does not exist at all.

\section{The Vacuum Energy Density}
Traditionally, in quantum field theory, all states that do not
contain net particle are assumed to be the same one. In
particular, the vacuum is assumed to be identical to the ground
state of the quantized field, while the ground state of the
quantized field is considered to be definite and unique, and it is
taken as $\otimes \prod_{\mathbf{k}}|+\frac{1}{2}\hbar
\omega_{\mathbf{k}}\rangle$, here $|+\frac{1}{2}\hbar
\omega_{\mathbf{k}}\rangle$ denotes the lowest energy state of
quantized field with frequency $\omega_{\mathbf{k}}$. This
directly or indirectly leads to series of difficulties in the
quantum field theory. The first and also the most obvious one is
the infinity of vacuum energy density by considering that there
are infinite number of vibration freedoms in the quantized field
and each has the lowest energy $+\frac{1}{2}\hbar
\omega_{\mathbf{k}}$. Undoubtedly, this is unreasonable.

The present work shows, if the negative energy field as well as
the positive one is also considered in the theory, the infinity of
the vacuum energy density can be immediately removed. Further, as
stated in the final section, by carefully distinguishing various
states which all have no net particle, the present work may supply
a possible way which does not violate any mathematical rules to
remove other infinities emerging in the quantum field theory.

Here, we take the electromagnetic field as an example in the
discussions, but the results can be easily generalized to other
fields. From Eq.~(\ref{Max-HamP}), the total Hamiltonian
consisting of both the positive and negative energy field is
\begin{eqnarray}\label{Max-Ham}
\mathcal{H} &=& \mathcal{H}^{(+)}+\mathcal{H}^{(-)}
\nonumber\\
&=& \sum_{r\mathbf{k}}\hbar\omega_{\mathbf{k}}
\zeta_{r}\left[a_{r}^{(+)\dag}(\mathbf{k})a^{(+)}_{r}(\mathbf{k})
- a_{r}^{(-)\dag}(\mathbf{k})a^{(-)}_{r}(\mathbf{k}) \right].
\end{eqnarray}
Since there is no net particle in the vacuum, one immediately gets
that the vacuum expectation of $\mathcal{H}$ must be 0. Thus, the
total energy density of vacuum is 0, not infinity as given in the
traditional quantum field theory.

\section{The Lamb-Shift}
It is well-known that, in terms of the Dirac theory, the
$2S_{1/2}$ and $2P_{1/2}$ levels of hydrogen are degenerate. In
1947, the measurements by Lamb and Retherford gave about $1000MHz$
for the level splitting $E(2S_{1/2})- E(2P_{1/2})$\cite{Lamb}.
This shift of the bound-state energy levels and the resulting
splitting are known as the Lamb-Shift.

Bethe in 1947 gave an approximate non-relativistic derivation of
the Lamb-Shift, obtaining a surprising good result considering the
nature of the calculation\cite{Bethe}.

Here the Lamb-Shift is recalculated in terms of the idea presented
in this work. It is shown that, by considering both positive and
negative energy field, the correct level shifts can be derived in
a very simple and direct way. On the one hand, the classic
non-relativistic second-order perturbation method is employed in
the derivation as Bethe had done; On the other hand, Bethe's
calculation is mainly based on the idea of mass renormalization of
electron, however, the present calculation does not concern the
renormalization process at all. It is shown that the same result
as Bethe's can still be obtained. The natural unit is taken in
this subsection for simplicity.

The interaction Hamiltonian of matter and positive and negative
energy field is
\begin{equation}\label{Int-Ham}
H_I^{(\pm)} = -\frac{e}{m}\mathbf{A}^{(\pm)}\cdot\mathbf{p}.
\end{equation}
Based on the present idea, we must consider the contributions of
both positive and negative energy field. First, for the positive
field, the level shift of a hydrogenic state $|n\ell\rangle =
\phi_{n\ell}(\mathbf{x})$ (where $n$ and $\ell$ are the principal
and angular momentum quantum number) is given by
\begin{eqnarray}\label{Lamb-P}
\delta E^{(+)}(n\ell) &=&
-\sum_{\lambda}\sum_{k}\sum_{r=1,2}\frac{|\langle\lambda,
n_{r}(\mathbf{k})=1|H_I^{(+)}|n\ell\rangle|^2}{E_{\lambda}+k-E_{n}}
\nonumber\\
&=&
-\sum_{\lambda}\sum_{k}\sum_{r=1,2}\left(\frac{e}{m}\right)^{2}
\frac{1}{2Vk}\frac{|\langle\lambda|\mathbf{\varepsilon}_{r}
(\mathbf{k})\cdot\mathbf{p}|n\ell\rangle|^2}{E_{\lambda}+k-E_{n}}\nonumber\\
&=&
-\frac{1}{6\pi^2}\left(\frac{e}{m}\right)^{2}\int_{0}^{\infty}kdk
\sum_{\lambda}\frac{|\langle\lambda|\mathbf{p}|n\ell\rangle|^2}{E_{\lambda}+k-E_{n}},
\end{eqnarray}
here the intermediate state $|\lambda, n_{r}(\mathbf{k})=1\rangle$
consists of the hydrogen atom in one of the complete set of states
$|\lambda\rangle \equiv \phi_{\lambda}(\mathbf{x})$ together with
one transverse photon, and $E_{\lambda}$ and $E_{n}$ are the
energy eigenvalues of $|\lambda\rangle$ and $|n\ell\rangle$; in
addition, $\langle\lambda|\mathbf{p}|n\ell\rangle = \int
d^3(\mathbf{x})\phi_{\lambda}^{*}(\mathbf{x})(-i\nabla)\phi_{n\ell}(\mathbf{x})$.

Next, the contributions of negative energy field to the level
shift of a hydrogenic state $|n\ell\rangle$ can be similarly
derived
\begin{equation}\label{Lamb-N}
\delta E^{(-)}(n\ell) =
-\frac{1}{6\pi^2}\left(\frac{e}{m}\right)^{2}\int_{0}^{\infty}kdk\sum_{\lambda}
\frac{|\langle\lambda|\mathbf{p}|n\ell\rangle|^2}{E_{\lambda}-k-E_{n}},
\end{equation}
here, virtual photon with negative energy $(-k)$ has been taken in
the calculations.

From Eq.~(\ref{Lamb-P}) and ~(\ref{Lamb-N}), the total level shift
is
\begin{eqnarray}\label{Lamb-Tot}
\delta E(n\ell) &=& \delta E^{(+)}(n\ell)+\delta E^{(-)}(n\ell)
\nonumber\\&=&
-\frac{1}{6\pi^2}\left(\frac{e}{m}\right)^{2}\int_{0}^{\infty}kdk\sum_{\lambda}
\left[\frac{|\langle\lambda|\mathbf{p}|n\ell\rangle|^2}{E_{\lambda}+k-E_{n}}+
\frac{|\langle\lambda|\mathbf{p}|n\ell\rangle|^2}{E_{\lambda}-k-E_{n}}\right]
\nonumber\\&=&
-\frac{1}{6\pi^2}\left(\frac{e}{m}\right)^{2}\int_{0}^{\infty}kdk\sum_{\lambda}
|\langle\lambda|\mathbf{p}|n\ell\rangle|^2\frac{2(E_{\lambda}-E_{n})}{(E_{\lambda}-E_{n})^2-k^2}
\nonumber\\&=&
\frac{1}{6\pi^2}\left(\frac{e}{m}\right)^{2}\sum_{\lambda}
|\langle\lambda|\mathbf{p}|n\ell\rangle|^2(E_{\lambda}-E_{n})\int_{0}^{\infty}\frac{2kdk}{k^2-(E_{\lambda}-E_{n})^2}.
\end{eqnarray}
Obviously, the integral in Eq.~(\ref{Lamb-Tot}) is logarithmic
divergent. By taking a cutoff value $K_{\lambda n}\sim
\sqrt{m(E_{\lambda}-E_{n})}$, one gets
\begin{equation}\label{Lamb-Tot2}
\int_{0}^{K_{\lambda
n}}\frac{2kdk}{k^2-(E_{\lambda}-E_{n})^2}=\ln\frac{m}{(E_{\lambda}-E_{n})}.
\end{equation}
In Bethe's calculation, he took $K_{\lambda n} = m$. Now, from
Eq.~(\ref{Lamb-Tot}) and (\ref{Lamb-Tot2}), the total level shift
is finally given by
\begin{equation}\label{Lamb-Tot3}
\delta
E(n\ell)=\frac{1}{6\pi^2}\left(\frac{e}{m}\right)^{2}\sum_{\lambda}
|\langle\lambda|\mathbf{p}|n\ell\rangle|^2(E_{\lambda}-E_{n})\ln\frac{m}{(E_{\lambda}-E_{n})}.
\end{equation}
The present result is exactly equal to Bethe's one\cite{Bethe}.

One notes that, in the derivation of Eq.~(\ref{Lamb-Tot3}), no
concepts such as the mass renormalization of electron and the
radiative corrections of electron self-energy, etc., are employed,
yet the same result as Bethe's is still obtained. Therefore, the
method presented here suggests and supplies us another possible
way for understanding the Lamb-Shift, maybe a more natural way in
the mathematics and physics.

\section{Particle and Wave}
E. Schr\"{o}dinger in 1920s suggested that the wavefunction
represents real entities, and a narrow wave packet just stands for
a particle\cite{Schrodinger}. He encountered a stumbling block, in
accordance with general mathematical laws, a wave packet
describing an isolated microscopic particle would rapidly
disperse, yet real particles are obviously more stable. Instead,
the Copenhagen interpretation of quantum mechanics, advocated by
N.Bohr and W.Henserberg et al. regards the wavefunction as merely
a mathematical tool representing an observer's subjective
knowledge of the system. According to Copenhagen's point of view,
quantum mechanics does not yield a description of an objective
reality but deals with probabilities of measurement outcomes. This
leads to a dualism between the wavefunction and the quantum
events. How do these quantum events arise? This interpretation
refused to go further, and deemed that question such as ``where
was the particle before I measured its position?'' is meaningless.

Dissatisfied with the doctrine of Copenhagen, de Broglie and D.
Bohm proposed that the particle is a localized and indivisble
entity; the wavefunction describes also a physically real field
which guides the movement of particle; without regard to
measurement the particle's location is always well
defined\cite{Bohm}. The criticism is if the wavefunction stands
for a real field and the pure wave theory itself is satisfactory,
then the associated particle seems to be superfluous. And what
exactly is this physically real wave field?

Thus, though quantum mechanics has made many remarkable
achievements, it is still shrouded a veil of mystery. To remove
agnosticism and mysticism, and further better understand the
reality of the wave function, one must probe deep into the essence
of matter waves, and clarify the relations between particle and
wave.

According to the present idea, the matter wave consists of both
positive and negative energy field, and the two fields correspond
to particles with completely opposite intrinsic properties. For a
wave field in pure state, if it consists of the same number of
positive and negative energy particles, then no net particle can
be observed; whereas if it contains one more positive energy
particles than negative ones, this then corresponds to single net
particle in the viewpoint of experimental observation, and the
corresponding wave field is the so-called `single particle wave'.
For such a `single particle wave', each positive energy particle
within the wave field has a probability to become the observed
particle in the measurement, though there is one and only one of
them which can become the observed particle in the measurement.
Specifically, if the wave field is described by wave function
$\psi(\mathbf{x}, t)$, then $|\psi(\mathbf{x}, t)|^2$ gives the
relative probability density that the positive energy particle at
$(\mathbf{x}, t)$ is observed in a measurement. In particular, for
the wave field described by $\psi(\mathbf{x}, t) =
\exp[i(\mathbf{k}\cdot\mathbf{x}-\omega t)]$, all positive energy
particles within the field have the same probability to be
observed in a measurement. On the other hand, if the particle at
$(\mathbf{x}, t)$ has been observed in measurement, the rest wave
field consists of equal number of positive and negative energy
particles, and then has no direct observation effect.

Further, the present idea manifests that the so-called `wave
packet reduction' known in the traditional quantum theory is
fictional. In line with the viewpoints mentioned above, once the
particle at $(\mathbf{x}, t)$ within the wave field has been
measured, its coupling with the original wave field is then
destroyed, and the rest wave field has then no direct observation
effect since it now consists of just equal number of positive and
negative energy particles. In this sense, the wave packet does not
disappear out of nothing, and it just becomes unobserved directly
after the measurement.

Not like Schr\"{o}dinger, according to the present viewpoint, a
wave packet is obviously different from any particle within the
wave field. The wave packet just tells the region into which
particles may be observed with relatively larger probabilities in
the measurement. The `motion' of the wave packet also does not
mean any particle's motion at all, and it just tells the change of
the most probable region along with the time. Similarly, the
`expansion' of the wave packet has nothing to do with any
particle's `dispersion' in the wave field.

Also not like de Broglie and D. Bohm, the present viewpoint shows
that the concept of a particle's trajectory is completely useless,
by considering what varies with the time is the oscillating state
of the entire wave field and not the placements of any particles
within the wave field. In this sense, the present idea is more
close to the Copenhagen interpretation than the de-Broglie and
Bohm's.

\section{Coherent Superposition: Double-Slit Interference, Entanglement}
As pointed out in previous section, if a wave field in pure state
consists of equal number of positive and negative energy particles
with opposite intrinsic properties, it has no direct observation
effect in measurements; however, this does not mean that such a
wave field has nothing to do with any experimental results at all.
Actually, many interesting phenomena and puzzling problems in
quantum theory arise from such pure wave fields.

For convenience in the following statements, we call a wave field
described by $|\psi\rangle$ the ``the recessive wave'' if the wave
field consists of equal number of positive and negative energy
particles, while call a wave field described by $|\psi\rangle$
``the dominant wave'' if the wave field contains more number of
positive energy particles than that of negative energy particles.
Though the recessive wave alone cannot be observed in experiment,
it can still manifest itself by coherently superposing with a
dominant wave.

As typical examples, let's firstly consider the well-known
double-slit interference. Specifically, for each individual
emission or receiving, the wave field is described by
\begin{equation}\label{Super-doub}
|\psi\rangle = \frac{1}{\sqrt{2}}\left(\mid\psi_{1}\rangle +
\mid\psi_{2}\rangle\right),
\end{equation}
here, $|\psi_{1}\rangle$ and $|\psi_{2}\rangle$ denote the waves
passing through slit 1 and 2, respectively. Either
$|\psi_{1}\rangle$ or $|\psi_{2}\rangle$ is the dominant wave,
while another one the recessive wave. Unless it is watched on, we
do not know which one on earth, $|\psi_{1}\rangle$ or
$|\psi_{2}\rangle$, is the dominant wave. The interference pattern
on the receiving screen originates from coherent superposition of
the recessive wave and dominant wave. On the one hand, for each
individual emission, there exists one and only one dominant wave
which passes through one of the two slits, and on the other hand,
there indeed exist two waves which simultaneously pass slit 1 and
2, respectively. Considering that there is one and only one
dominant wave, any experiments designed for watching on which way
the particle chooses must tell one and only slit through which the
particle passes; at the same time, once the coupling between the
particle (which has been observed) and the wave field is cut off
due to experimental observation, the expected interference pattern
is then destroyed.

The same idea can also be applied to entangled systems. Consider a
pair of particles with spin-$1/2$ that are in a state in which the
total spin is zero. In such experiments, such two particles can be
produced by a single particle decay. They separate, and after a
long time no longer interact. On the hypothesis, that are not
disturbed, the law of angular momentum conservation guarantees
that they remain in a singlet state.

Taking the total spins and its $z$ component as quantum numbers,
the singlet state may be written in the form
\begin{equation}\label{Super-entang}
|\psi\rangle =
\frac{1}{\sqrt{2}}\left(\mid\uparrow\rangle_{1}\mid\downarrow\rangle_{2}
- \mid\downarrow\rangle_{1}\mid\uparrow\rangle_{2}\right),
\end{equation}
where the subscripts refer to the particles. As is well known, the
singlet state (\ref{Super-entang}) is an entangled state and
cannot be factorized. In recent years, theoretical and
experimental works seem to show that entangled systems may violate
the separability principle\cite{Buscemi,Christensen,Popescu}.
However, how are the `nonlocal correlations' between entangled
systems established? The present work insists that, between
entangled systems, there indeed exists a kind of quantum
correlation that is different from any classical correlations,
however, no superluminal signal transmits between them at all.

The basic idea is: either
$\mid\uparrow\rangle_{1}\mid\downarrow\rangle_{2}$ or $
\mid\downarrow\rangle_{1}\mid\uparrow\rangle_{2}$ is the dominant
wave, and another one the recessive wave, for each individually
prepared pair of particles. Due to the existence of the recessive
component (in Eq.(\ref{Super-entang})), and its coherent
superposition with the dominant wave, the correlations between two
particles thus become nonclassical, as demonstrated in some
correlation experiments which are frequently investigated in
recent years. In this sense, the present point of view is
obviously different from the hidden variable theory, and closer to
the orthodox point of view. On the other hand, in terms of the
Copenhagen interpretation, the roles of two components of
$\mid\psi\rangle$ are completely equal in status, and neither of
them is in a subordinate position. Therefore, the orthodox
viewpoint insists that the spin value of each particle is
indeterminate before measuring, and then if the spin of one
particle is determined in a local measuring, the state of another
one immediately reduces to the corresponding component with
opposite spin value, no matter how far from they are.

In the light of the present idea, the roles of two terms in
entangled state (\ref{Super-entang}) are not equal: one and only
one of them is the dominant wave which consists of two more
positive energy particles than negative ones, and it is this
dominant component that definitely specifies what the spin value
is for either local measurement. For instance, for a given pair of
entangled particles described by Eq. (\ref{Super-entang}),
assuming that $\mid\uparrow\rangle_{1}\mid\downarrow\rangle_{2}$
($\mid\downarrow\rangle_{1}\mid\uparrow\rangle_{2}$) is the
dominant (recessive) component, then the result of a local
measuring along $z$ direction for particle 1 must tell
$s_{1z}=+\frac{1}{2}\hbar$. More importantly, in accordance with
the scheme proposed here, the value of $s_{2z}$ is independent of
any local measuring on the particle 1. The result can obviously be
applied to any other directions by considering that the state
vector describing entangled state Eq. (\ref{Super-entang}) has the
same form in different basis.

\section{Discussions: Vacuum versus The Ground State}

As shown previously, to remove the infinity of vacuum energy
density, one needs no more knowledge about the property of vacuum,
but that there is no net particle in the vacuum. On the other
hand, in the traditional quantum field theory, it is taken for
granted that, $|V\rangle = |G\rangle = \otimes
\prod_{\mathbf{k}}\left|+\frac{1}{2}\hbar
\omega_{\mathbf{k}}\right\rangle$, here $|V\rangle$ and
$|G\rangle$ denote the vacuum and ground state of the quantized
field, respectively, and this means that all the lowest energy
states of the field are realized with probability $100\%$ in the
vacuum, and further the infinite vacuum energy density then
arises.

However, in the light of the present idea, we need consider both
positive and negative energy field. On the one hand, the energy
density of the vacuum equals 0 as pointed out previously,, and on
the other hand, just like $+\frac{1}{2}\hbar \omega_{\mathbf{k}}$
is higher than $0$ for positive energy field, $-\frac{1}{2}\hbar
\omega_{\mathbf{k}}$ is also higher than $0$ for negative energy
field for each $\omega_{\mathbf{k}}$, thus deviations from $0$ to
$\pm\frac{1}{2}\hbar \omega_{\mathbf{k}}$ for positive and
negative energy field must become increasingly difficult with
increasing of $\omega_{\mathbf{k}}$. Without loss of generality,
let $Q(\omega_{\mathbf{k}})$ be the probability with which the
lowest energy states $|+\frac{1}{2}\hbar
\omega_{\mathbf{k}}\rangle$ and $|-\frac{1}{2}\hbar
\omega_{\mathbf{k}}\rangle$ of the positive and negative energy
field are realized in the vacuum. A natural supposition is:
$Q(\omega_{\mathbf{k}})\rightarrow 0$ as
$\omega_{\mathbf{k}}\rightarrow +\infty$, and
$Q(\omega_{\mathbf{k}})\rightarrow 1$ as
$\omega_{\mathbf{k}}\rightarrow 0$. The requirements can be easily
fulfilled by taking
\begin{equation}\label{GS-Prob}
Q(\omega_{\mathbf{k}}) = \exp\left(-\frac{\hbar
\omega_{\mathbf{k}}}{2k_{B}T_{0}}\right),
\end{equation}
here, $k_B$ is the Boltzmann constant, and $T_0$ a formal
temperature owned to the vacuum.

First, if $T_0 = +\infty$, from Eq.~(\ref{GS-Prob}),
$Q(\omega_{\mathbf{k}}) = 1$ for $\forall\omega_{\mathbf{k}}\in
(0, +\infty)$. Actually, this is just the choice implied in the
traditional quantum field theory. In other words, the traditional
quantum field theory implies that the vacuum has an infinite
formal temperature.

Second, if $T_0 = 0$, from Eq.~(\ref{GS-Prob}),
$Q(\omega_{\mathbf{k}}) = 0$ for $\forall\omega_{\mathbf{k}}\in
(0, +\infty)$. This means that no any ground state of the field
can be realized at all in the vacuum, and this further means that
all second and higher order corrections based on perturbation
method be zero.

Finally, for $0<T_0<+\infty$, one gets
$0<Q(\omega_{\mathbf{k}})<\infty$ for
$\forall\omega_{\mathbf{k}}\in (0, +\infty)$.

Discarding the above mentioned two extreme cases, i.e., $T_0 =
+\infty$ and $T_0 = 0$, by using Eq.~(\ref{GS-Prob}), the energy
density of the positive energy field in the vacuum can then be
calculated,
\begin{equation}\label{Vau-Den}
\rho_{0}^{(+)} = \frac{\hbar
c}{(2\pi)^2}\left(\frac{2k_BT_0}{\hbar c}\right)^{4}\Gamma(4).
\end{equation}
On the other hand, the vacuum energy density of the negative
energy field $\rho_{0}^{(-)} = -\rho_{0}^{(+)}$. For instance, for
$T_0 = 3K$, one gets $\rho_{0}^{(+)} = 0.225 \times
10^{-12}J/m^3$.

How do we determine that a state with no net particle is the
vacuum state or not in the theory? First, if a particle is created
from or annihilated to the vacuum state, it must be a virtual
particle. Second, considering that a matter system cannot directly
exchange energy, momentum, or any other observed quantities, with
the vacuum, a virtual particle accompanying with the vacuum state
must not change the state of an isolated matter system, i.e., the
initial and final state must be the same state. A state which
fulfill both requirements is the vacuum state.

According to the present idea, the ground state of the field is
different from the vacuum state; in the vacuum state, neither net
particle nor net field exist, whereas in the ground state,
although no net particle but the positive energy field still
exists in principle.

In traditional quantum field theory, the ground state of the field
is assumed to be unique, and taken as:
$\otimes\prod_{\mathbf{k}}\left|+\frac{1}{2}\hbar
\omega_{\mathbf{k}}\right\rangle$. This leads to infinite energy
for the ground state. However, if the negative energy field is
considered in the theory, it is found that the above assumption
about the ground state is completely unnecessary, and the infinity
resulting from this assumption is thus unreal.

A more natural assumption about the ground state is: for pure
state $|\psi\rangle = \sum_{\mathbf{k}}c(\mathbf{k})\left|\hbar
\omega_{\mathbf{k}}+\frac{1}{2}\hbar
\omega_{\mathbf{k}}\right\rangle$, the corresponding ground state
is $|G\rangle_{\psi} =
\sum_{\mathbf{k}}c(\mathbf{k})\left|+\frac{1}{2}\hbar
\omega_{\mathbf{k}}\right\rangle$, here $|c(\mathbf{k})|^2=1$.

\end{document}